\UseRawInputEncoding
\documentclass{jaa}

\usepackage{graphicx}
\usepackage{url}
\usepackage[colorlinks,linkcolor=blue,urlcolor=blue]{hyperref}
\usepackage{xcolor}
\usepackage{array}
\usepackage{booktabs}
\usepackage{pdflscape}
\usepackage[normalem]{ulem}
\newcommand{\et}{  {\it et al.}}
\newcommand\farcdeg{\mbox{$.\!\!^\circ$}}%
\newcommand\farcsec{\mbox{$.\!\!^{\prime\prime}$}}%

\newcommand{\aap}{    {\it A \& A}}

\newcommand{\actaa}{  {\it Acta Astron.}}

\newcommand{\aj}{     {\it AJ}}
\newcommand{\apj}{    {\it ApJ}}
\newcommand{\apjl}{   {\it ApJL}}

\newcommand{\mnras}{  {\it MNRAS}}
\newcommand{\na}{     {\it New Astronomy}}
\newcommand{\nat}{    {\it Nature}}
\newcommand{\pasp}{   {\it Pub. Astron. Soc. Pac.}}

\begin{document}\sloppy

\title{UOCS. IV. Discovery of diverse hot companions to blue stragglers in the old open cluster King 2
}


\author{Vikrant V. Jadhav\textsuperscript{1,2,*}, Sindhu Pandey\textsuperscript{3}, Annapurni Subramaniam\textsuperscript{1} \and Ram Sagar\textsuperscript{1,3}}
\affilOne{\textsuperscript{1}Indian Institute of Astrophysics, Sarjapur Road, Koramangala, Bangalore, India.\\}
\affilTwo{\textsuperscript{2}Joint Astronomy Programme and Department of Physics, Indian Institute of Science, Bangalore, India.\\}
\affilThree{\textsuperscript{3}Aryabhatta Research Institute of Observational Sciences, Manora Peak, Nainital, India\\}


\twocolumn[{

\maketitle

\corres{vikrant.jadhav[at]iiap.res.in}

\msinfo{31 August 2020}{31 August 2020}

\begin{abstract}
King 2, one of the oldest clusters in the Milky Way, with an age of $\sim$ 6 Gyr and distance of $\sim 5700$ pc, has been observed with UVIT payload on the \textit{ASTROSAT}. With membership information derived from {\it Gaia} EDR3, the cluster is found to have 39 blue straggler stars (BSSs). We created multi-wavelength spectra-energy distributions (SED) of all the BSSs. Out of 10 UV detected BSSs, 6 bright ones fitted with double component SEDs and were found to have hotter companions with properties similar to extreme horizontal branch (EHB)/subdwarf B (sdB) stars, with a range in luminosity and temperature, suggesting a diversity among the hot companions. We suggest that at least 15\% of BSSs in this cluster are formed via mass-transfer pathway. When we compared their properties to EHBs and hotter companions to BSS in open and globular clusters, we suggest that EHB/sdBs like companions can form in binaries of open clusters as young as 6 Gyr. 
\end{abstract}

}]


\doinum{}
\artcitid{\#\#\#\#}
\volnum{000}
\year{0000}
\pgrange{1--}
\setcounter{page}{1}
\lp{1}

\section{Introduction} \label{sec:intro}

The evolution of binary systems strongly depends on the initial orbital parameters and its further evolution, where any change in their orbits can lead to a widely different evolution. If one of the stars evolves and fills its Roche lobe, the system will undergo mass transfer. Details such as duration and rate of mass transfer will depend on the orbits and masses of the binary stars. If such a binary is present in a star cluster, and the secondary of the binary has mass similar to the main sequence turnoff (MSTO) mass, then the secondary will become brighter than the MSTO and appear as a blue straggler star (BSS). Otherwise, the secondary will be fainter than the MSTO and can be classified as a blue lurker. The blue lurkers are identified by their high stellar rotation (Liener \et\ 2019) or evidence of extremely low-mass (ELM) white dwarf (WD) companion (Jadhav \et\ 2019). 
Depending on the evolutionary status of binary components, the binary system can be observed as main sequence (MS)+MS, contact binaries (Rucinski 1998), common envelop, MS+horizontal branch (HB; Subramaniam \et\ 2016), MS+extreme HB (EHB; Singh \et\ 2020), MS+ subdwarf-B (sdB; Han \et\ 2002), MS+WD (Jadhav \et\ 2019), WD+WD (Marsh \et\ 1995) and many more combinations. The binary evolution also depends on external factors such as collisions in a high-density environment which can decouple the binary (Heggie 1995) and tertiary star which can expedite the mass transfer/merger by reducing the orbital separation (Kozai 1962). 

We are carrying out a long term project\footnote{UOCS: UVIT Open Cluster Study} of characterising products of binary stars such as BSSs in open clusters (OCs). Ultraviolet imaging of binary systems reveals the presence of hotter companions in the binary system, given that the hotter companion is luminous in UV. Old OCs such as NGC 188 and NGC 2682 are rich with BSSs, binary stars and contain many such optically sub-luminous UV-bright companions (Subramaniam \et\ 2016; Sindhu \et\ 2019; Jadhav \et\ 2019). Similar companions have been identified to BSSs in the outskirts of GCs (Sahu \et\ 2019; Singh \et\ 2020). Subramaniam \et\ (2020) provides a summary of the BSSs and post mass transfer systems in star clusters.

\begin{table}
    \centering
    \caption{Age, distance, reddening (E(B$-$V)) and metallicity of King 2 estimated by various investigators are listed.}
    \begin{tabular}{cc ccr}
    \toprule
	Age	&	Distance	&	E(B$-$V)	&	Metallicity	&	Ref.	\\
	(Gyr)	&	(pc)	&	(mag)	&		&		\\ \hline
	6.02	&	5750	&	0.31	&	-0.42	&	[1]	\\
	6	&	5690$\pm$65	&	0.31$\pm$0.02	&	-0.5 to -2.2	&	[2]	\\
	& & &	-0.32	&	[3]	\\
 4 to 6	&$\sim 7000$&0.23 to 0.5 &	&	[4]	\\	
\bottomrule
\multicolumn{5}{l}{[1] Dias {\it et al.} (2002), [2] Aparicio {\it et al.} (1990), } \\
\multicolumn{5}{l}{[3] Tadross (2001), [4] Kaluzny (1989)}		
    \end{tabular}
      \label{tab:info}
\end{table}

King 2 is one of the oldest clusters in the Milky Way, with an age of $\sim$ 6 Gyr and distance of $\sim 5700$ pc (Table~\ref{tab:info}). However, it has been poorly studied due to its considerable distance and unknown membership information. 
For identifying and characterising hot BSSs and their possible companions, we obtained Ultra Violet Imaging Telescope (UVIT)/\textit{ASTROSAT} observations of rich OC King 2 ($\alpha_{2000} = 12\farcdeg{75};\ \delta_{2000} = + 58\farcdeg{183};\ l = 122\farcdeg{9}$ and $b = - 4\farcdeg{7}$) under ASTROSAT proposal A02\_170.
Kaluzny (1989) presented the first optical colour-magnitude diagram (CMD) study of this distant cluster using \textit{BV} CCD photometric data. This yielded a range of plausible ages and distances for different assumed reddenings and metallicities. The galactocentric distance of the cluster was estimated to be $\sim$ 14 Kpc. Aparicio \et\ (1990) (A90 hereafter) did a comprehensive study on the cluster using \textit{UBVR} photometry and derived an age of 6 Gyr and a distance of 5.7 kpc for solar metallicity. They also indicated the presence of a good fraction of binaries in the MS. Tadross (2001) estimated a value of [Fe/H] $= - 0.32$ using the ($U - B$) colour excess from the literature data, while Warren \& Cole (2009; WC09 hereafter) derived a value of [Fe/H] $= - 0.42\pm0.09$ using spectroscopic data. These metallicity estimates are significantly sub-solar and inconsistent with the finding of A90. WC09 found a distance of 6.5 kpc and a slightly younger age, $\sim$ 4Gyr, better fitted the optical CMD and 2MASS Ks, red clump if the reddening is adopted as E(B$-$V) = 0.31 mag. This distance puts King 2 at RGC = 13 kpc, where its metallicity falls close to the trend of the galactic abundance gradients derived in Friel \et\ (2002). There has been no proper motion study available for this cluster till {\it Gaia} DR2 (Gaia Collaboration \et\ 2018). 
Cantat-Gaudin \et\ (2018) provided a membership catalogue of King 2 with 128 members with \textit{Gaia} DR2, and Jadhav \et\ (2021) provided kinematic membership of 1072 stars (and 340 probable members) using kinematic data taken from {\it Gaia} EDR3.

Above discussed optical photometric studies indicate a good number of post-MS hot stars in King 2. In fact, Ahumada \et\ (2007) have identified 30 BSS candidates based on the location of these stars in the cluster. We present the UVIT and the archival data used in this study in the next section, followed by analyses, results and discussion.

\section{UVIT and archival data}
\label{sec:data}

\begin{figure}
    \centering
    \includegraphics[width=0.47\textwidth]{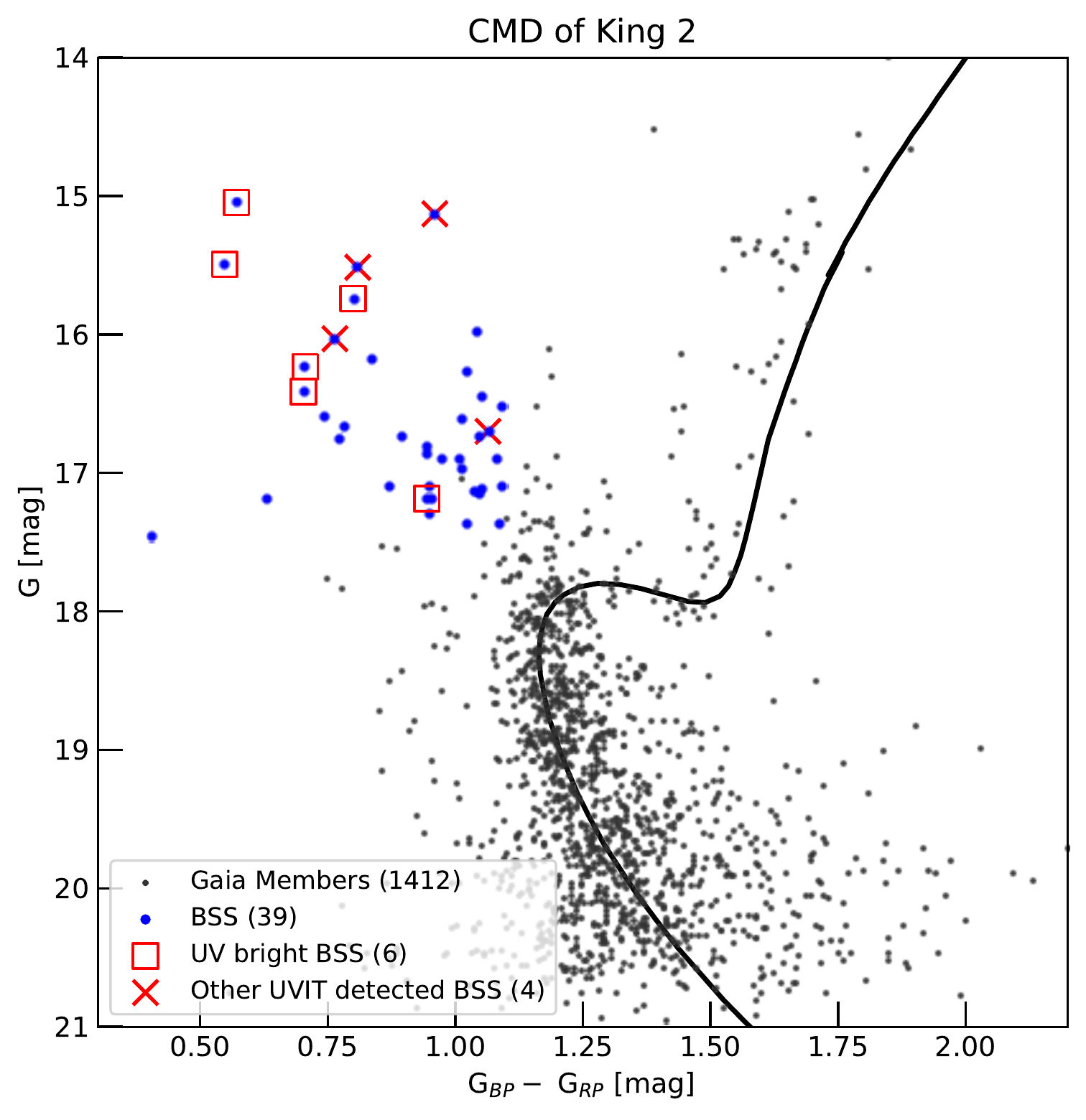}
    \caption{CMD of King 2 cluster candidates using {\it Gaia} EDR3 data. 
    All BSS members are shown as blue circles. The UV bright BSS (see section. \ref{sec:sed_fitting}) are shown as red squares, and other UVIT detected BSS are shown as red X's.
    The {\it Gaia} members are shown as black dots along with the PARSEC isochrone of log $age$ = 9.7, [M/H] = -0.4, DM = 13.8 and E(B$-$V) = 0.45.}
    \label{fig:CMD}
\end{figure}

We observed King 2 with UVIT, which is one of the payloads on the first Indian multiwavelength space observatory \textit{ASTROSAT}, launched on 28 September 2015. The observation was carried out by UVIT on 17 December 2016, simultaneously in two filters. The telescope has three channels with a set of filters in them: Far-UV (FUV; 130 - 180 nm), near-UV (NUV; 200 - 300 nm) and visible (VIS; 350 - 550 nm), where the VIS channel is intended to correct the drift of the spacecraft (see Kumar \et\ 2012 and Tandon \et\ 2017 for more details). The cluster was observed in one FUV (F148W, limiting magnitude $\approx$ 23 mag) and one NUV (N219M, limiting magnitude $\approx$ 22 mag) filter for exposure time of $\sim$2.7 ksec. The FWHM of PSF in F148W and F219N images is 1\farcsec33 and 1\farcsec35 respectively.
The data reduction was done using {\sc ccdlab} (Postma \& Leahy 2017) and PSF photometry was performed using {\sc daophot} package of {\sc iraf} (Tody 1993). More details of the reduction process are presented in Jadhav {\it et al.} (2021). We have detected ten member stars in either F148W and/or N219M filter. 

\begin{figure*}
    \centering
    \includegraphics[width = 0.9\textwidth]{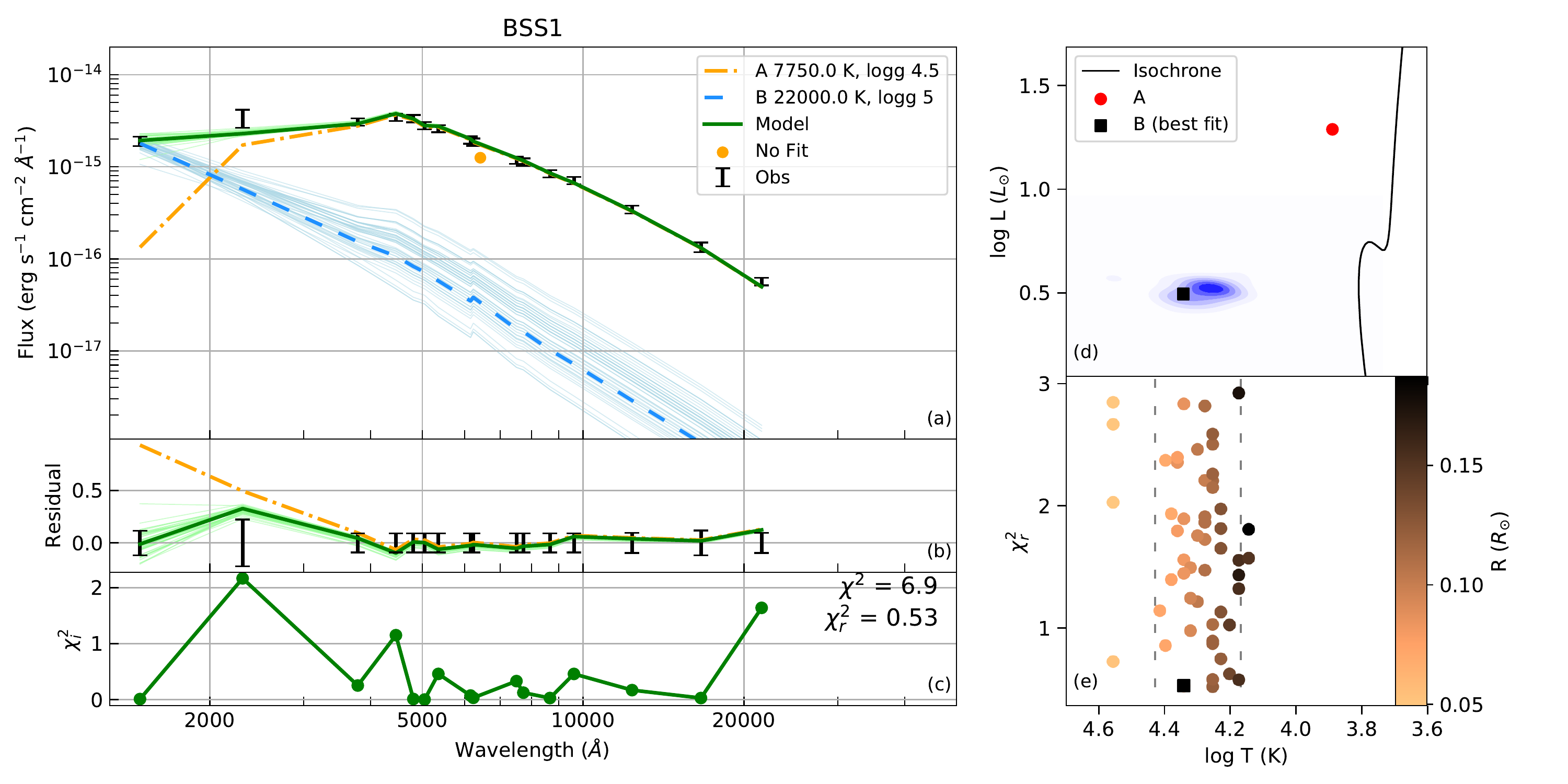}
    \caption{Two-component SED of BSS1. (a) Composite SED (green curve) is shown along with the observed flux (as black error-bars). The unfitted point (in this case: CAHA.R) is shown as orange dot. The cooler (BSS, orange dot-dashed curve) and hotter (blue dashed curve) component are also shown with their T$_{eff}$ and log $g$. The model, B component and residuals of noisy iterations are also shown as light coloured lines. (b) The fractional residual is shown for single component fit (orange dot-dashed curve) and composite fit (green solid curve). The fractional observational errors are also indicated on X-axis. (c) The $\chi_i^2$ of each data point. (d) H--R diagram of the two components along with the isochrone for reference. The density distribution of the noisy B component fits is plotted in blue. (e) T$_{eff}$--$\chi^2$ distribution for the noisy B component fits coloured according to their radii. The dashed lines are the quoted limits of the temperature.}
    \label{fig:SEDs_Kurucz_1}
\end{figure*} 

We obtained archival optical (\textit{UBVR}) photometry data from A90 catalogue (Calar Alto observatory; CAHA) and cross-matched with UVIT data using {\sc topcat} (Taylor 2005). The cluster was observed with \textit{GALEX} under All-sky Imaging (AIS) survey in NUV filter (exp. time $\sim$ 100 sec).
All the detected member stars were further cross-matched with photometric data from UV to IR wavelength bands obtained from \textit{GALEX} (Bianchi \et\ 2000), PAN-STARRS PS1 (Chambers \et\ 2016), {\it Gaia} EDR3 (Gaia Collaboration \et\ 2020), 2MASS (Skrutskie \et\ 2006), WISE (Wright \et\ 2010) using virtual observatory tools in VOSA (Bayo \et\ 2008).

\section{SED fitting and Colour Magnitude diagrams} \label{sec:sed_fitting}

The data were corrected for reddening (E(B$-$V) = 0.31$\pm$0.02) using Fitzpatrick (1999) and Indebetouw \et\ (2005) and calibrated with the cluster distance of 5750$\pm$100 pc (we have overestimated the error to cover distance estimates from Dias \et\ (2002) and A90). We have adopted the metallicity of [Fe/H]= $-$0.5 for all the stars and used Kurucz model spectrum (Castelli \et\ 1997) for comparison. The SED fitting was done as follows:
\begin{enumerate}
    \item We constructed the observed spectral energy distribution (SED) for all stars using the data from UV to IR wavelength, as mentioned above. The fluxes of all stars are given in Table~\ref{tab:flux_table}.
    \item Kurucz models (Castelli \et\ 1997) of log $g \in (3.0,5.0)$ were fitted to optical and IR points (above 3000 \AA) using VOSA\footnote{\url{http://svo2.cab.inta-csic.es/theory/vosa/index.php}} (Bayo \et\ 2008). There were some sources which showed UV excess in multiple UV points compared to the model fit. We selected such stars to be fitted with a two component SED. Otherwise, the single component fits are deemed satisfactory and are stated in the lower part of Table~\ref{tab:parameters}.
    \item We used the cooler component parameters from above fits and then fitted a hotter component to the residual using \texttt{Binary\_SED\_Fitting}\footnote{\url{https://github.com/jikrant3/Binary_SED_Fitting} \texttt{Binary\_SED\_Fitting} is a python code which uses $\chi^2$ minimisation technique to fit two component SEDs.}. In preliminary double component fits, the hotter components were found to be compact objects, hence they were fitted with log $g = 5$ Kurucz models.
    \item Very small errors in PAN-STARRS PS1, \textit{Gaia} EDR3 and A90 photometry led to ignoring relatively high error UV data-points, hence they were replaced with mean errors for better residual across all wavelengths. A few data points were removed to achieve better fits and lower $\chi^2$ (see Fig.~\ref{fig:SEDs_Kurucz_1} (a)). The error values given in Table~\ref{tab:flux_table} are original unmodified errors. The detailed SED fitting method is explained in section 3.3 of Jadhav \et\ (2019).
    \item The best fit parameters for single stars or cooler components are taken from the VOSA fits. The hotter component parameters are taken from the least $\chi^2$ model in the two component fitting. 
    \item The errors in cooler component parameters are fairly low and are taken as the grid values. To derive errors in the hotter component parameters, we used a statistical approach. We first generated 100 iterations of observed SEDs with added Gaussian noise in each data point. These 100 SEDs were then fitted with double components. However, not all the double fits converged, hence we only kept hotter components with $6000 K < T_{eff} < 37000 K$. Logarithmic distributions of the parameters from the noisy \& converging iterations were then fitted with Gaussian distributions. FWHM of these Gaussian distributions are defined as the upper and lower limits of the fitting parameters (temperature, radius and luminosity).
\end{enumerate}



\begin{table*}
    \centering
    \footnotesize
    \begin{tabular}{lccccc cc r}
    \toprule
    \multicolumn{9}{c}{Double Fits}\\ 
Name	&	Comp.	&	log $g$	&	Teff			&	R			&	L			&	Scaling Factor	&	N$_{fit}$	&	$\chi_r^2$	\\
	&		&		&	[K]			&	[$R_{\odot}$]			&	[$L_{\odot}$]			&		&		&	($\chi_{r,single}^2$)	\\ \hline
BSS1	&	A	&	4.5	&	7750	$\pm$	125	&	2.44	$\pm$	0.04	&	19.3	$\pm$	1.8	&	8.99E-23	&	16	&	0.5 (10.5)	\\
	&	B	&	5	&	22000	$^{+4842}	_{-7269}$	&	0.122	$^{+0.150}	_{-0.037}$	&	3.1	$^{+0.7}	_{-0.4}$	&	2.23E-25	&		&		\\ 
BSS2	&	A	&	3.5	&	8250	$\pm$	125	&	3.72	$\pm$	0.06	&	57.6	$\pm$	3.4	&	2.09E-22	&	16	&	3.9 (1.2)	\\
	&	B	&	5	&	24000	$^{+6802}	_{-4996}$	&	0.234	$^{+0.131}	_{-0.089}$	&	16.4	$^{+1.5}	_{-1.2}$	&	8.27E-25	&		&		\\ 
BSS3	&	A	&	4.5	&	7250	$\pm$	125	&	3.56	$\pm$	0.06	&	31.6	$\pm$	2.1	&	1.92E-22	&	16	&	1.0 (16.1)	\\
	&	B	&	5	&	24000	$^{+3229}	_{-10186}$	&	0.094	$^{+0.191}	_{-0.016}$	&	2.7	$^{+0.8}	_{-0.3}$	&	1.34E-25	&		&		\\ 
BSS4	&	A	&	5	&	8000	$\pm$	125	&	2.16	$\pm$	0.04	&	17.2	$\pm$	1.5	&	7.06E-23	&	17	&	2.3 (9.6)	\\
	&	B	&	5	&	26000	$^{+1912}	_{-13323}$	&	0.089	$^{+0.304}	_{-0.011}$	&	3.3	$^{+0.8}	_{-0.3}$	&	1.20E-25	&		&		\\ 
BSS5	&	A	&	3.5	&	6500	$\pm$	125	&	2.29	$\pm$	0.04	&	8.4	$\pm$	0.8	&	7.90E-23	&	12	&	0.4 (19.9)	\\
	&	B	&	5	&	14000	$^{+12479}	_{-5493}$	&	0.237	$^{+0.376}	_{-0.150}$	&	1.9	$^{+2.1}	_{-0.5}$	&	8.47E-25	&		&		\\ 
BSS7	&	A	&	3	&	8500	$\pm$	125	&	2.90	$\pm$	0.05	&	39.6	$\pm$	1.4	&	1.27E-22	&	11	&	0.6 (1.4)	\\
	&	B	&	5	&	19000	$^{+8088}	_{-2901}$	&	0.270	$^{+0.102}	_{-0.134}$	&	8.6	$^{+1.2}	_{-0.8}$	&	1.11E-24	&		&		\\

\bottomrule
\end{tabular}
\begin{tabular}{lcccc ccccc r}
\multicolumn{11}{c}{Single Fits}\\
Name	&	log $g$	&	Teff	&	e\_Teff	&	R	&	e\_R	&	L	&	e\_L	&	Scaling Factor	&	N$_{fit}$	&	$\chi_r^2$	\\
	&		&	[K]	&	[K]	&	[$R_{\odot}$]	&	[$R_{\odot}$]	&	[$L_{\odot}$]	&	[$L_{\odot}$]	&		&		&		\\ \hline
BSS6	&	4	&	6500	&	125	&	2.32	&	0.04	&	8.74	&	0.41	&	8.15E-23	&	11	&	3.4	\\
BSS8	&	3	&	7000	&	125	&	4.19	&	0.07	&	38.17	&	2.05	&	2.66E-22	&	11	&	3.2	\\
BSS9	&	3.5	&	7500	&	125	&	2.81	&	0.05	&	22.58	&	1.14	&	1.20E-22	&	11	&	4.2	\\
BSS10	&	3	&	6250	&	125	&	2.91	&	0.05	&	11.67	&	0.54	&	1.28E-22	&	11	&	3.9	\\
BSS11	&	4	&	6750	&	125	&	5.28	&	0.09	&	52.55	&	2.64	&	4.22E-22	&	11	&	1.6	\\
BSS12	&	3.5	&	7000	&	125	&	1.97	&	0.03	&	8.66	&	0.43	&	5.85E-23	&	12	&	47.9	\\
BSS13	&	3.5	&	6750	&	125	&	2.32	&	0.04	&	10.17	&	0.51	&	8.16E-23	&	11	&	3.2	\\
BSS14	&	3	&	6750	&	125	&	2.38	&	0.04	&	10.63	&	0.48	&	8.54E-23	&	11	&	2.5	\\
BSS15	&	3	&	6500	&	125	&	2.44	&	0.04	&	9.71	&	0.49	&	9.02E-23	&	11	&	3.6	\\
BSS16	&	3	&	6750	&	125	&	1.99	&	0.03	&	7.42	&	0.37	&	6.00E-23	&	11	&	3.1	\\
BSS17	&	3.5	&	7000	&	125	&	2.32	&	0.04	&	11.60	&	0.55	&	8.12E-23	&	11	&	4.3	\\
BSS18	&	4	&	6500	&	125	&	2.22	&	0.04	&	8.08	&	0.34	&	7.48E-23	&	15	&	12.3	\\
BSS19	&	4.5	&	7250	&	125	&	2.89	&	0.05	&	20.92	&	1.17	&	1.26E-22	&	8	&	13.3	\\
BSS20	&	5	&	6500	&	125	&	2.23	&	0.04	&	8.05	&	0.32	&	7.53E-23	&	11	&	3.4	\\
BSS21	&	5	&	6500	&	125	&	3.08	&	0.05	&	15.32	&	0.61	&	1.43E-22	&	11	&	4.9	\\
BSS22	&	3.5	&	6250	&	125	&	3.28	&	0.06	&	14.88	&	0.62	&	1.63E-22	&	11	&	1.5	\\
BSS23	&	3.5	&	6500	&	125	&	2.84	&	0.05	&	13.00	&	0.59	&	1.22E-22	&	11	&	3.7	\\
BSS24	&	3	&	7250	&	125	&	2.12	&	0.04	&	11.24	&	0.59	&	6.77E-23	&	11	&	2.1	\\
BSS25	&	5	&	6250	&	125	&	2.64	&	0.05	&	9.73	&	0.39	&	1.06E-22	&	15	&	11.7	\\
BSS26	&	3	&	7250	&	125	&	2.28	&	0.04	&	13.06	&	0.67	&	7.89E-23	&	15	&	11.3	\\
BSS27	&	4.5	&	6250	&	125	&	2.17	&	0.04	&	6.47	&	0.26	&	7.10E-23	&	11	&	4.3	\\
BSS28	&	4	&	7250	&	125	&	2.23	&	0.04	&	12.57	&	0.68	&	7.54E-23	&	15	&	9.8	\\
BSS29	&	4	&	6500	&	130	&	2.27	&	0.04	&	8.40	&	0.41	&	7.81E-23	&	5	&	125.1	\\
BSS30	&	4.5	&	6250	&	125	&	2.45	&	0.04	&	8.35	&	0.34	&	9.08E-23	&	11	&	5.3	\\
BSS31	&	5	&	8250	&	127	&	1.40	&	0.02	&	8.09	&	0.35	&	2.98E-23	&	15	&	9.0	\\
BSS32	&	4	&	6250	&	125	&	2.57	&	0.04	&	9.27	&	0.39	&	9.98E-23	&	15	&	19.5	\\
BSS33	&	4	&	6500	&	125	&	2.04	&	0.04	&	6.79	&	0.28	&	6.29E-23	&	15	&	10.9	\\
BSS34	&	4	&	6500	&	125	&	3.80	&	0.07	&	23.42	&	1.01	&	2.19E-22	&	11	&	4.2	\\
BSS35	&	3	&	6250	&	125	&	2.41	&	0.04	&	8.07	&	0.38	&	8.75E-23	&	11	&	4.4	\\
BSS36	&	3.5	&	6500	&	125	&	2.37	&	0.04	&	9.13	&	0.44	&	8.51E-23	&	11	&	3.3	\\
BSS37	&	5	&	6500	&	125	&	2.25	&	0.04	&	8.23	&	0.34	&	7.65E-23	&	11	&	3.6	\\
BSS38	&	5	&	6500	&	125	&	2.68	&	0.05	&	11.70	&	0.48	&	1.09E-22	&	11	&	3.4	\\
BSS39	&	3.5	&	6500	&	125	&	3.31	&	0.06	&	17.67	&	0.80	&	1.65E-22	&	11	&	3.9	\\

\bottomrule
    \end{tabular}
    \caption{Fitting parameters of the best fit of the double and single component fits of BSSs with the hotter component. Scaling factor is the value by which the model has to be multiplied to fit the data, N$_{fit}$ is the number of data points fitted and $\chi_r^2$ is the reduced $\chi^2$ for the composite fit. The $\chi_r^2$ values of single fits of the cooler components are given in brackets. \textit{ Note: the log $g$ values are imprecise due to the insensitivity of the SED to log $g$.}}
    \label{tab:parameters}
\end{table*}

Fig.~\ref{fig:CMD} shows the CMD of 1412 cluster candidates identified from {\it Gaia} EDR3 with probability of over 50\% (Jadhav \et\ 2021) and are marked as grey points.
Among these stars, we have selected 39 member stars brighter ($G<17.5$ mag) and bluer ($G_{BP}-G_{RP} < 1.1$ mag) than the MSTO as BSSs.
Seven of them were detected in F148W, and seven were detected in N219M (four in both filters). An isochrone of log $age$ $=$ 9.7 is over-plotted on the CMD. 20 of the BSSs are detected in the NUV filter of \textit{GALEX}. We fitted Kurucz model SEDs to all BSSs and found excess UV flux in 15 BSS (BSS 1, 2, 3, 4, 5, 7, 8, 9, 10, 19, 26, 28, 29, 33 and 36). Among these, BSS1, 2, 3, 4, 5 and 7 have multiple UV data points from UVIT or \textit{GALEX} or both. Only these six were fitted with double component SEDs, because a hot component fit can be reliable if the number of UV data points is more than one. Hereafter, these six BSS will be referred to as `UV bright BSSs' and others will be referred to as `UV faint BSSs'. The four BSS detected in UVIT but not fitted with hotter component are shown as red X's in the CMD.

We have shown an example of a double component SED fit of BSS1 in Fig.~\ref{fig:SEDs_Kurucz_1} (a).
The BSS1-A component is a BSS with 7750 K, while the BSS1-B component has T$_{eff}$ of 22000 K. The reduction in residual after including the hotter component is visible in Fig.~\ref{fig:SEDs_Kurucz_1} (b). The $\chi_i^2$ for individual points is shown in Fig.~\ref{fig:SEDs_Kurucz_1} (c) with $\chi_r^2$ of 0.53. Although, we note that the $\chi_r^2$ need not be $\sim$ 1, for a non-linear model fitting (Andrae \et\ 2010). One has to look at residuals and $\chi^2$ both to determine the goodness of fit.
Fig.~\ref{fig:SEDs_Kurucz_1} (d) shows the Hertzsprung--Russell (H--R) diagram of A and B components. The density distribution of noisy \& converging iterations is also shown to get an idea of degeneracy in temperature and luminosity. Fig.~\ref{fig:SEDs_Kurucz_1} (e) panel shows the best fit and the noisy \& converging iterations in T$_{eff}$--$\chi^2$ phase-plane.
The double component fits of BSS2, 3, 4, 5 and 7, and single-component fits of BSS10 and BSS15 are shown in Fig.~\ref{fig:SEDs_Kurucz_2}. The fitting parameters are mentioned in Table~\ref{tab:parameters}.

\begin{figure}
    \centering
    \includegraphics[width=0.48\textwidth]{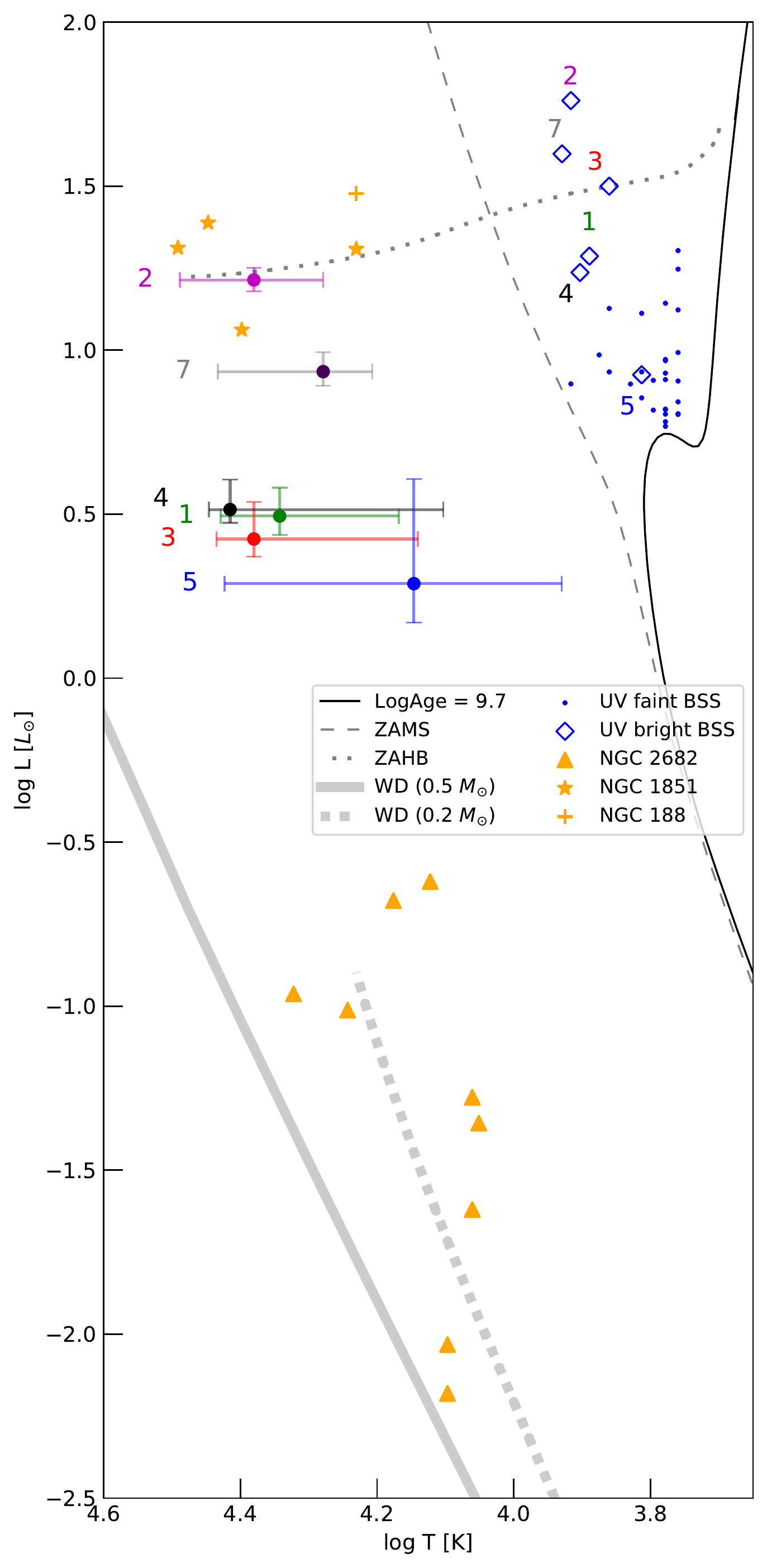}
    \caption{H--R diagram of locations of components of binaries in King 2, NGC 2682 (Sindhu \et\ 2019; Jadhav \et\ 2019; Sindhu \et\ in prep.), NGC 188 (Subramaniam \et\ 2016) and EHB stars in NGC 1851 (Singh \et\ 2020) . In King 2, the UV faint BSS (blue dots), UV bright BSS (blue diamonds with their ID) and hotter components in BSSs (coloured filled circles with numbers and error bars) are shown. Hotter components in NGC 2682 and NGC 188 are shown as orange triangles and cross. EHB stars in NGC 1851 are shown as stars.
    The PARSEC isochrone (black curve), PARSEC zero age MS (dashed grey line), WD cooling curves (thick grey curves) and BaSTI ZAHB (dotted grey curve) are shown for reference.}
    \label{fig:comparison_CMD}
\end{figure}

The H--R diagram of the BSSs detected in King 2 is shown in Fig.~\ref{fig:comparison_CMD}.
We have shown the UV faint BSSs as blue dots, UV bright BSSs are represented as blue diamonds, and the hotter components of UV bright BSSs as filler circles. 
We have taken the parameters of the hotter companions detected along with the BSSs in NGC 188 from Subramaniam \et\ (2016) and NGC 2682 from Sindhu \et\ (2019), Jadhav \et\ (2019) and Sindhu \et\ (in prep). They are plotted in the figure as orange cross and triangles respectively.
The parameters of the EHB stars in NGC 1851 are taken from Singh \et\ (2020) and are shown as orange stars. 
The PARSEC\footnote{\url{stev.oapd.inaf.it/cgi-bin/cmd}} isochrone of log $age$ $=$ 9.7 is over-plotted and shown in black (Bressan \et\ 2012), along with the WD cooling curves\footnote{\url{www.astro.umontreal.ca/~bergeron/CoolingModels/}} (thick grey curves, Tremblay \& Bergeron 2009) and BaSTI\footnote{\url{basti-iac.oa-abruzzo.inaf.it/isocs.html}} zero age HB (ZAHB; dashed black curve; Hidalgo \et\ 2018). 

\section{Results and Discussion} \label{sec:Discussion}

\textbf{BSS and their companions in literature:}

The BSSs have T$_{eff}$ range of 5750 to 8500 K and radii of 1.4 to 5.21 \(R_{\odot}\). By comparison to isochrones, they have mass in the range of 1.2 to 1.9 \(M_{\odot}\), the brightest BSS being 3 mag brighter than the MSTO. 
Majority of BSSs in King 2 have T$_{eff}$ similar to the older NGC 188 (6100--6800 K; Gosnell \et\ 2015), but are cooler than NGC 2682 (6250--9000 K; Sindhu \et\ in prep.), which is expected due to its slightly younger age.
The BSSs in NGC 188 (Geller \& Mathieu 2011; Gosnell \et\ 2014; Subramaniam \et\ 2016) and NGC 2682 (Sindhu \et\ 2019) are known to have evolved companions. The companions were classified as WD, ELM WDs, post-AGB/HB according to their luminosity and temperature. 

BSS2-A, BSS3-A and BSS7-A lie above/on the ZAHB in Fig.~\ref{fig:comparison_CMD}. There is a degeneracy in this region of the H--R diagram where one could find both massive BSS as well as ZAHB stars. Stars in these two evolutionary phases will have different masses (HB mass $<$ MSTO; BSS mass $>$ MSTO), that could be used to lift the degeneracy.
Bond \& Perry (1971) measured the masses of stars in this region of the NGC 2682 CMD and determined that they are indeed high mass BSSs. One star is found in this region of the NGC 188 CMD and it is classified as a BHB (Rani et al. 2020), this star is significantly brighter than the rest of the BSSs. In the case of King 2, the BSSs show a continuous distribution up to the brightest BSS, hence BSS2-A/BSS3-A are most likely normal BSSs. However, their mass estimations (via log $g$ measurements or asteroseismology) are required before confirming their evolutionary status.

\noindent \textbf{What are the hotter companions?}

The hotter companions in UV bright BSSs have T$_{eff}$ of 14000 to 26000 K (spectral type B) and radii of 0.09 to 0.27 \(R_{\odot}\). 
Fig.~\ref{fig:comparison_CMD} shows the density distributions of the best 100 fits for the hotter companions. Fig.~\ref{fig:comparison_CMD} also shows the location of various companions to BSS in NGC 188 and NGC 2682, and EHB stars in NGC 1851, one of which has a BSS as its companion (Singh \et\ 2020).  

The hotter companion to BSSs in NGC 2682 are all fainter and near the WD cooling curves. While those for NGC 1851 and NGC 188 lie closer to the ZAHB region. In King 2, the limiting magnitude of UVIT observations is 23 and 22 mag in F148W and N219M, respectively. According to WD cooling models (corrected for distance and extinction; Tremblay \& Bergeron 2009), only WDs younger than 0.7 and 16 Myr old would be detectable. As seen in Fig.~\ref{fig:comparison_CMD}, the hotter companions are well above the WD cooling curves, and these are not WDs.

The hotter companions are likely to be hot HB stars which are also known as EHB stars or sdB stars, as inferred from their T$_{eff}$, radii and luminosity. These are core-helium burning stars with T$_{eff}$ in the range of 20,000 - 40,000 K and are compact (0.15-0.35 R$_\odot$; Heber 2016; Sahoo \et\ 2020). As these stars are hot and not as small as WDs, they appear bright in the UV. These stars are thought to contribute to the UV upturn seen in elliptical or in early-type galaxies (Brown \et\ 1997). 
The sdB stars have very thin hydrogen envelope and are thought to be the stripped core of a red giant star (Heber 2016). Maxted \et\ (2001) found a good fraction of the sdB stars in detached, but short period binary systems. sdB stars are thought to provide important clues to common envelope evolution in tight binaries.

BSS2-B lies on the blue end of ZAHB and very similar to the EHBs in the outskirts of GC NGC 1851. Similarly, hot and bright post-AGB/HB candidate was found as a companion to a BSS in NGC 188 (Subramaniam \et\ 2016). Hence, BSS2-B could be an EHB star. 
BSS7-B is near a known EHB from NGC 1851 and is slightly fainter than ZAHB, hence, it is likely to be an EHB.
BSS5-B lies slightly above the WDs. Hence, it can be a very young He-WD (Panei \et\ 2007) or an sdB star.
Rest of the King 2 BSS companions are slightly fainter than the ZAHB, and they are likely sdBs.

\noindent \textbf{Formation pathways of BSSs and EHBs/sdBs:}

BSS formation mechanism involves mass gain while the EHB/sdB formation involves the stripping of the envelope of a post-MS star.
The detection of EHB/sdBs confirms `binary mass transfer' as the formation mechanism for BSS1 to BSS5 and BSS7. As the cooler companions are BSSs that are supposed to have gained mass, we can infer that the detected EHBs/sdBs have transferred mass to the BSSs companions. Therefore, the BSS+EHB/sdB systems in King 2 illustrate stars on both sides of the mass exchange. We see a range in their temperature and luminosity, suggesting a diversity among the hotter companions.

The lifespan of sdB stars is expected to be between 100 to 200 Myr (Bloemen \et\ 2014; Schindler \et\ 2015 and references therein), after which they descend the WD cooling curve. When the current BSS expands, the mass transfer is expected to start again due to the short orbit, which already allowed the previous instance of mass transfer. The system can begin stable/unstable mass transfer and become a WD+WD system. Alternatively, it can merge through a common-envelop phase and become a massive WD. The exact evolution will depend on then orbital parameters, mass transfer efficiency and mass loss. 

King 2 is one of the oldest OC, lies in the outskirts of the galactic disk. It is metal-poor compared to the Galactic disc OCs. The environment is quite similar to outskirts of GCs, which are also metal-poor, old and of comparable density. While most of the BSSs in GCs lie in core and are formed via mergers (Chatterjee \et\ 2013), BSSs in outskirts of GCs can form through mass transfer as seen in EHB-4 of NGC 1851 (Singh \et\ 2020). Our study suggests that the at least 15\% of the BSSs in King 2 are formed via mass transfer pathway of formation.

We have seen sdB companions to BSSs in NGC 188 and NGC 1851 (both are older systems), however none in NGC 2682 (which is younger). NGC 6791 of slightly younger age also has sdB stars (Kaluzny \& Udalski 1992; Reed \et\ 2012). This might suggest that there is an upper age limit of $\sim$5 Gyr for the formation of sdB stars in OCs.

\section{Conclusions and Summary} \label{sec:conclusion}
\begin{itemize}
    \item The old OC King 2 has a large population (39) of BSSs, spreading up to 3 mag brighter than the MSTO. We constructed SEDs of all the BSS using UV to IR data. The BSSs have T$_{eff}$ in the range of 5750--8250 K, luminosity in the range of 5.6--57.5 L$_{\odot}$ and mass in the range of 1.2--1.9 M$_{\odot}$.
    \item Six of the UV bright BSS showed excess UV flux and were successfully fitted with double component SEDs. The hotter components have T$_{eff}$ of 14000--26000 K and R/R$_{\odot}$ of 0.09--0.27, suggesting a range of properties. Two of the hotter companions to the BSS are likely EHB stars, while four are likely sdB stars.
    \item EHB/sdB companions imply that these 6 (out of 39) BSSs have formed via binary mass transfer. The SED fits show that sdB stars can be created in old OCs such as King 2 (similar to old OC NGC 188 and GC NGC 1851). 
\end{itemize}

Spectroscopic time series and radial velocity variations can uncover the binary nature as well as properties of these systems and help in estimating the log $g$ and mass of these stars. The mass and orbital estimations will expand our knowledge of BSS, EHB and sdB formation scenarios.

\section*{Acknowledgements}

We thank the anonymous referee for their valuable comments and inputs. RS would like to thank the National Academy of Sciences, India (NASI), Prayagraj for the award the NASI Honorary Scientist; Alexander von Humboldt Foundation, Germany for the award of long-term group research linkage program and Director, IIA for hosting and making available facilities of the institute.
UVIT project is a result of a collaboration between Indian Institute of Astrophysics (IIA), Bengaluru, The Inter-University Centre for Astronomy and Astrophysic (IUCAA), Pune, Tata Institute of Fundamental Research (TIFR), Mumbai, several centres of Indian Space Research Organisation (ISRO), and Canadian Space Agency (CSA). 
This work has made use of data from the European Space Agency (ESA) mission {\it Gaia} (\url{https://www.cosmos.esa.int/gaia}), processed by the {\it Gaia} Data Processing and Analysis Consortium (DPAC, \url{https://www.cosmos.esa.int/web/gaia/dpac/consortium}).
This publication makes use of VOSA, developed under the Spanish Virtual Observatory project supported by the Spanish MINECO through grant AyA2017-84089.
\vspace{-1em}



\appendix
\section{Single and composite SEDs of BSS}

\begin{figure*}
    \centering
    \includegraphics[width = 0.8\textwidth]{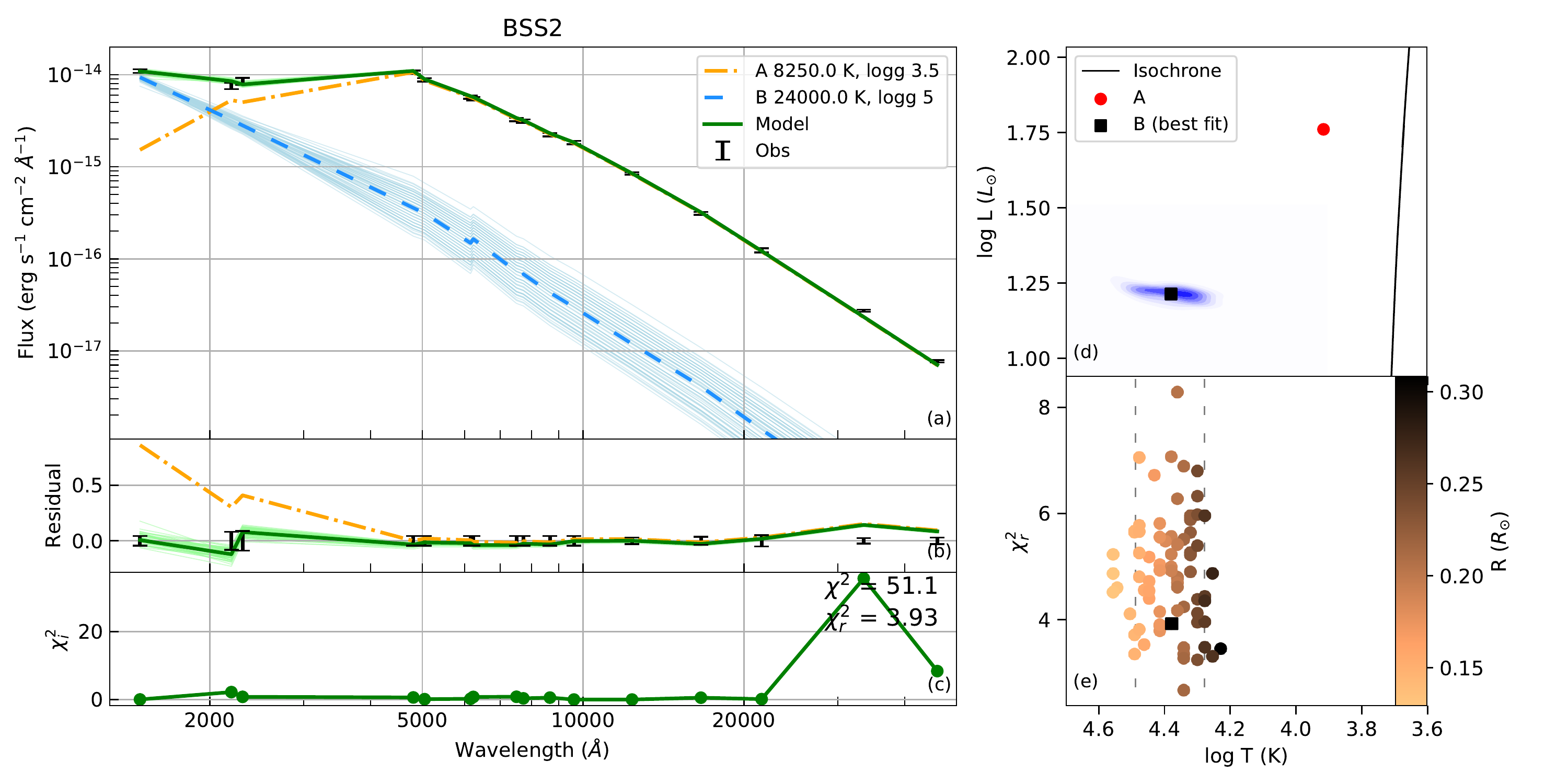}
    \includegraphics[width = 0.8\textwidth]{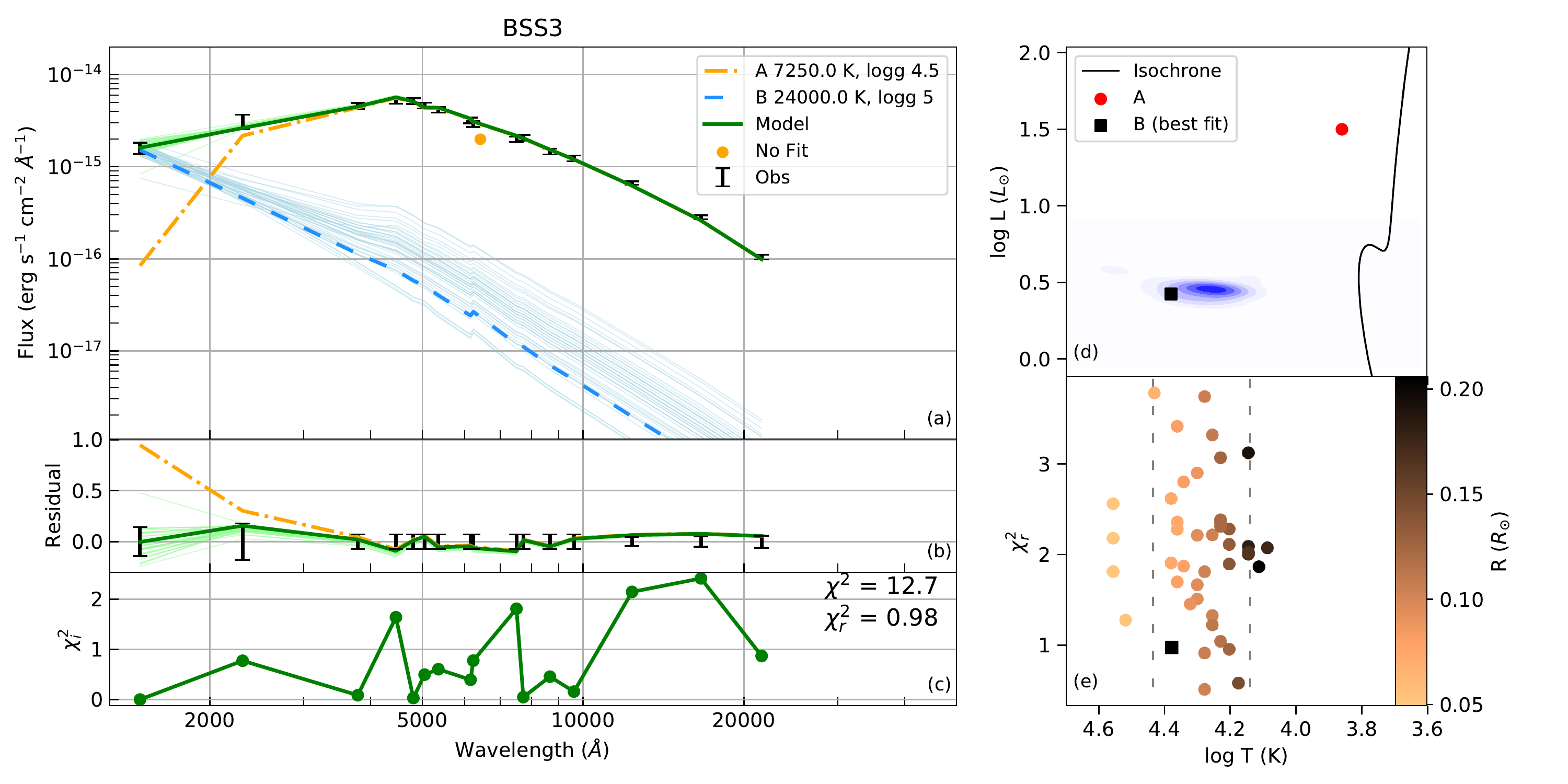}
    \includegraphics[width = 0.8\textwidth]{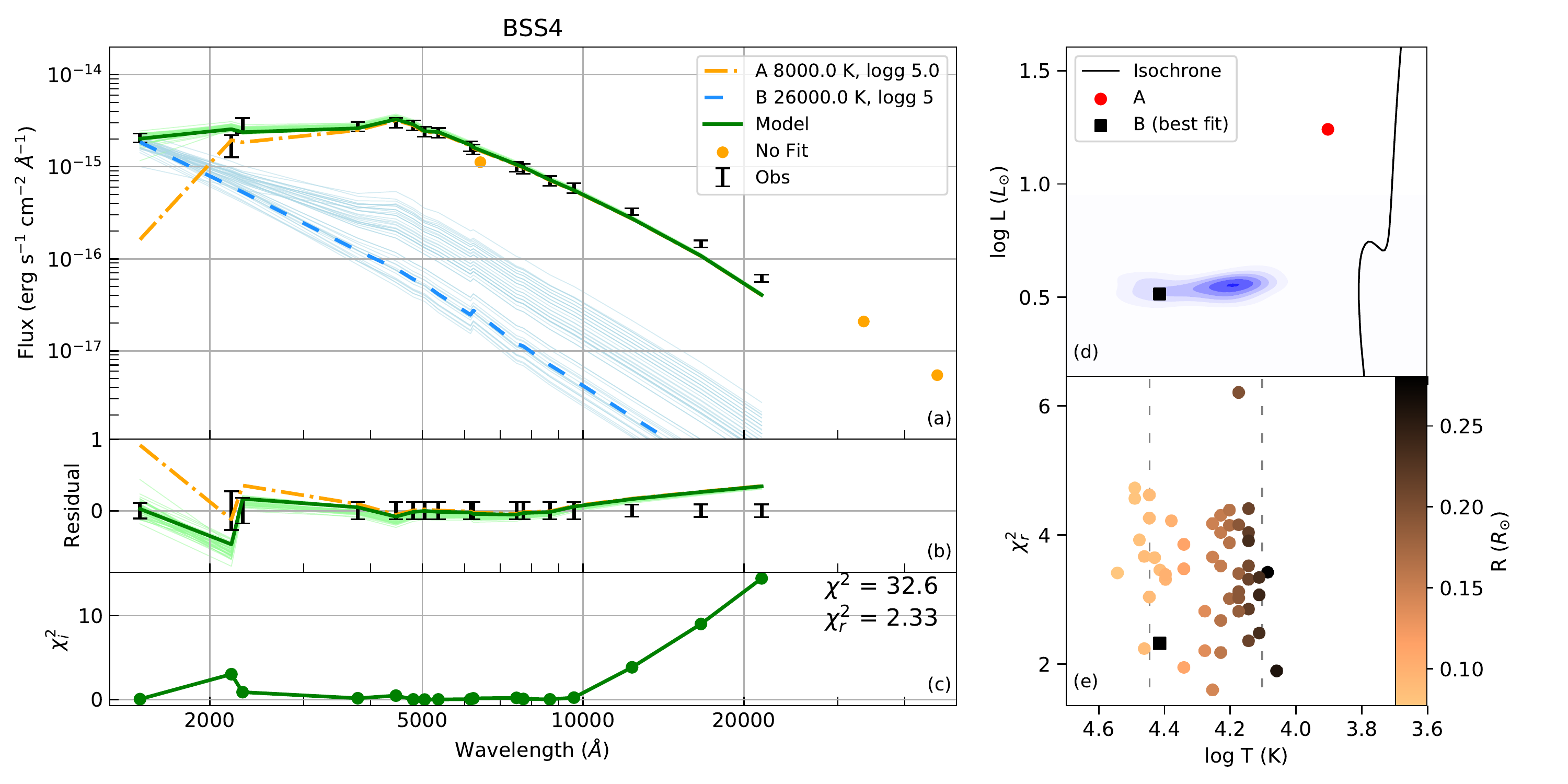}
    \caption{The descriptions of double component fits are same as Fig.~\ref{fig:SEDs_Kurucz_1}.}
    \label{fig:SEDs_Kurucz_2}
\end{figure*}

\renewcommand{\thefigure}{\arabic{figure} (Continued...)}
\addtocounter{figure}{-1}

\begin{figure*}
    \centering
    \includegraphics[width = 0.8\textwidth]{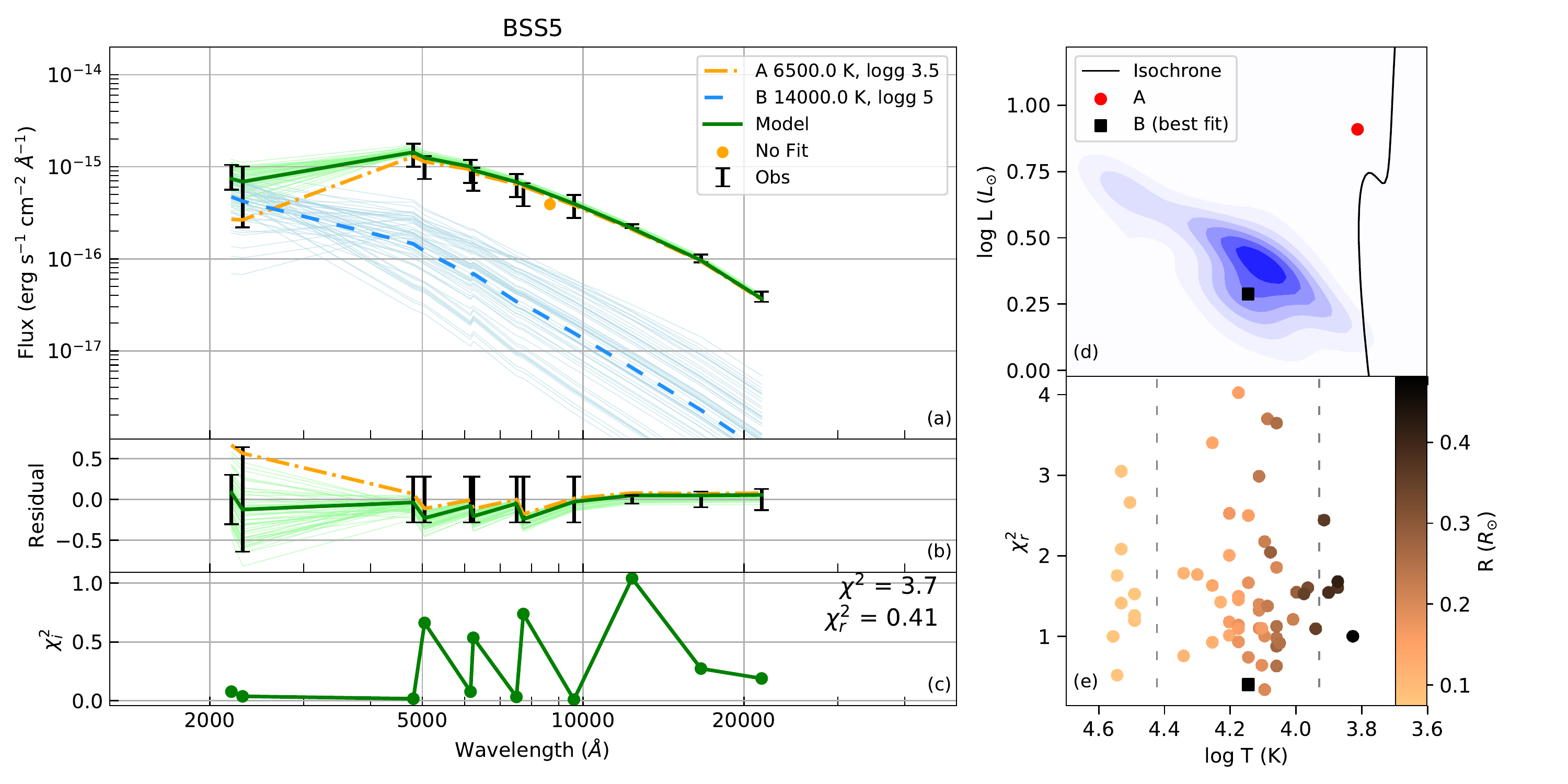}
    \includegraphics[width = 0.8\textwidth]{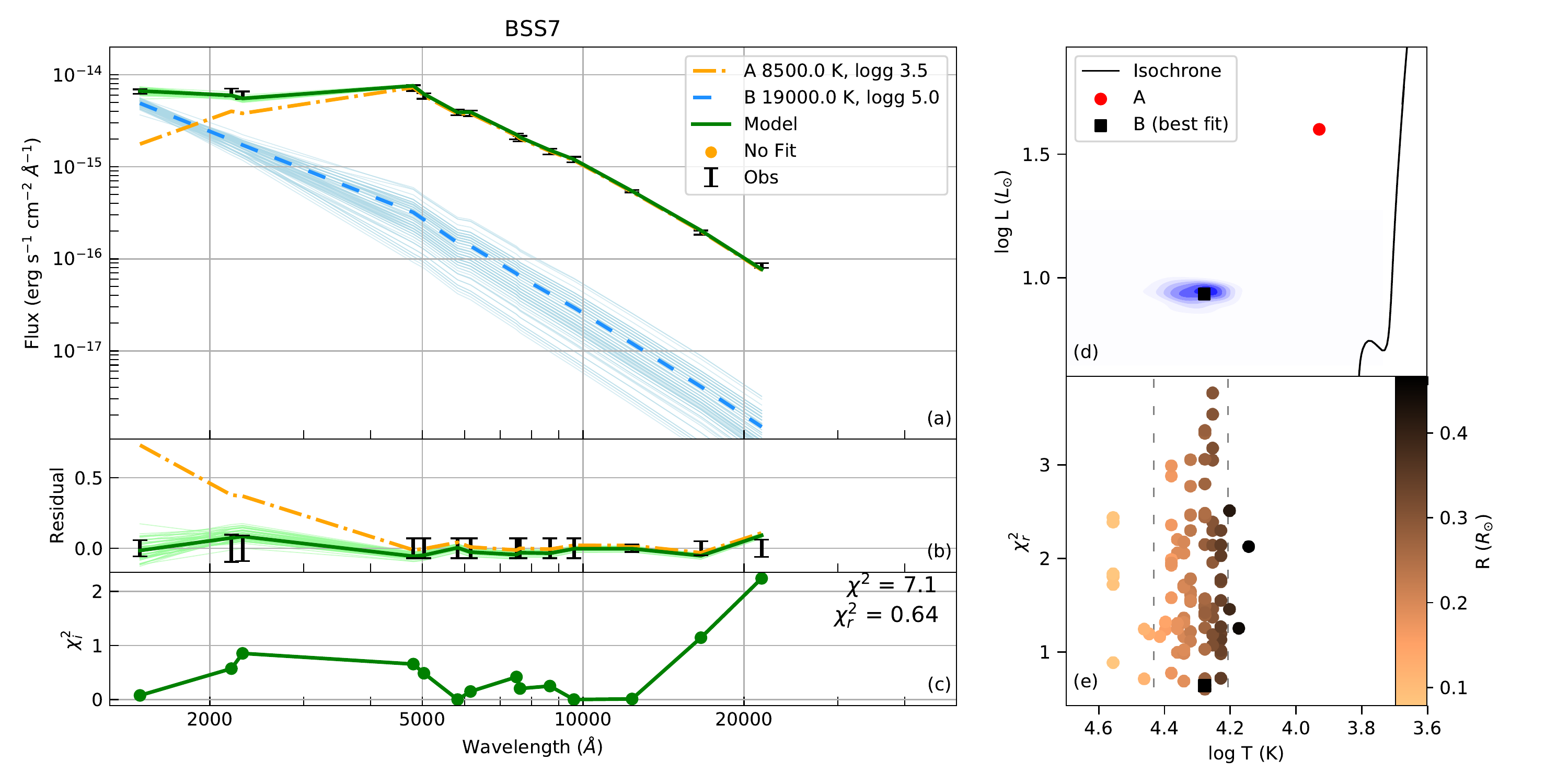}
    \includegraphics[width=0.49\textwidth]{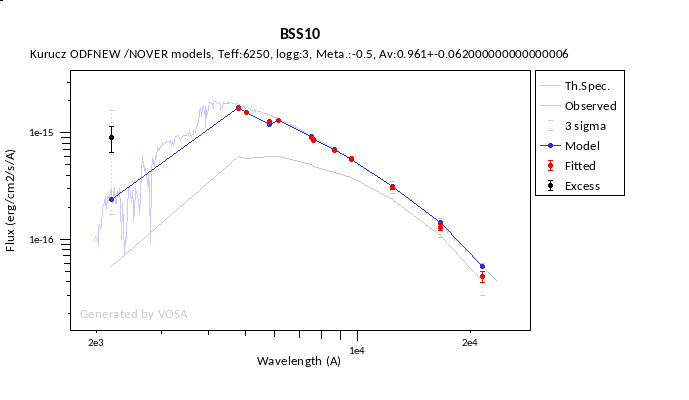}
    \includegraphics[width=0.49\textwidth]{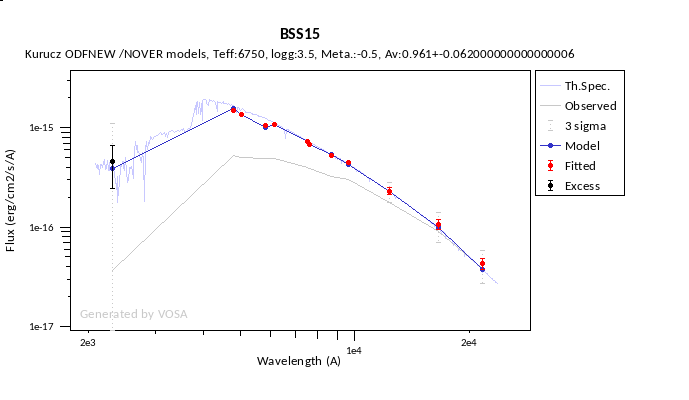}
    \caption{The single component fits of BSS10 and BSS15 are shown as an example with model fit (blue curve), fitted data points (red points) with 1 $\sigma$ and 3 $\sigma$ errors as solid and dashed lines. The theoretical spectra (in grey) is added for reference. The observed (reddening affected) SED is shown in grey below the corrected data-points. The title mentions the T$_{eff}$, log $g$, metallicity and A$_V$ of the model fit.
    }   
\end{figure*}

\renewcommand{\thefigure}{\arabic{figure}}

\begin{landscape}
\begin{table}
\caption{Coordinates, Gaia EDR3 source IDs, flux and flux errors of the stars in all used filters. All flux are given in [erg s$^{-1}$ cm$^{-2}$ \AA$^{-1}$]. GALEX, Gaia (EDR3), Pan-Starrs (PS1) and 2MASS photometry is taken from archives. CAHA photometry is taken from A90.}
\label{tab:flux_table}
\footnotesize
\begin{tabular}{lrrrlllll}
\toprule
Name   &        R.A (J2016) &        Dec. (J2016) &          \textit{Gaia} EDR3 source\_id &     UVIT.F148W$\pm$err &     UVIT.N219M$\pm$err &      GALEX.NUV$\pm$err &         CAHA.U$\pm$err &         CAHA.B$\pm$err \\
\midrule
BSS1  &  12.735411 &  58.185675 &  424416887106018688 &  1.87e-16$\pm$2.18e-17 &                    --- &  2.74e-16$\pm$6.08e-17 &  8.05e-16$\pm$1.29e-17 &  1.11e-15$\pm$1.45e-17 \\
BSS2  &  12.851678 &  58.152155 &  424415066039931520 &  1.08e-15$\pm$5.07e-17 &  4.67e-16$\pm$3.57e-17 &  6.86e-16$\pm$6.07e-17 &                    --- &                    --- \\
BSS3  &  12.748154 &  58.207693 &  424418398934484480 &  1.57e-16$\pm$2.16e-17 &                    --- &  2.51e-16$\pm$4.47e-17 &  1.21e-15$\pm$1.94e-17 &  1.68e-15$\pm$2.20e-17 \\
BSS4  &  12.741835 &  58.196669 &  424418364574755328 &  2.03e-16$\pm$2.17e-17 &   1.07e-16$\pm$2.9e-17 &  2.30e-16$\pm$4.14e-17 &  7.21e-16$\pm$9.45e-18 &  9.75e-16$\pm$9.03e-18 \\
BSS5  &  12.676827 &  58.114184 &  424416234271059072 &                    --- &  4.98e-17$\pm$1.53e-17 &  4.93e-17$\pm$3.16e-17 &                    --- &                    --- \\
BSS6  &  12.852115 &  58.202338 &  424415684515168768 &                    --- &                    --- &                    --- &                    --- &                    --- \\
BSS7  &  12.497494 &  58.135206 &  425074017100947968 &  6.42e-16$\pm$3.65e-17 &  3.96e-16$\pm$3.82e-17 &  4.86e-16$\pm$4.40e-17 &                    --- &                    --- \\
BSS8  &  12.475572 &  58.232940 &  425168433361005568 &  1.96e-16$\pm$2.09e-17 &   1.31e-16$\pm$2.8e-17 &                    --- &                    --- &                    --- \\
BSS9  &  12.562694 &  58.306732 &  425170361807707392 &  7.94e-17$\pm$1.43e-17 &                    --- &  1.76e-16$\pm$4.34e-17 &                    --- &                    --- \\
BSS10 &  12.951969 &  58.156780 &  424414207046475904 &                    --- &  5.57e-17$\pm$1.51e-17 &                    --- &                    --- &                    --- \\
BSS11 &  12.601162 &  58.118577 &  424323394258415744 &                    --- &  1.14e-16$\pm$2.18e-17 &  1.23e-16$\pm$2.83e-17 &                    --- &                    --- \\
BSS12 &  12.718310 &  58.184925 &  424416917160868480 &                    --- &                    --- &   8.45e-17$\pm$3.9e-17 &  3.00e-16$\pm$1.16e-17 &  4.42e-16$\pm$1.49e-17 \\
BSS13 &  12.668926 &  58.042374 &  424321809405171968 &                    --- &                    --- &  3.74e-17$\pm$1.55e-17 &                    --- &                    --- \\
BSS14 &  12.677840 &  58.058054 &  424322226027437440 &                    --- &                    --- &                    --- &                    --- &                    --- \\
BSS15 &  12.630396 &  58.055428 &  424322569624703744 &                    --- &                    --- &  3.68e-17$\pm$1.72e-17 &                    --- &                    --- \\
BSS16 &  12.578670 &  58.094139 &  424322874557099008 &                    --- &                    --- &                    --- &                    --- &                    --- \\
BSS17 &  12.637498 &  58.089431 &  424323016301297792 &                    --- &                    --- &  6.69e-17$\pm$3.16e-17 &                    --- &                    --- \\
BSS18 &  12.729170 &  58.183177 &  424416921465758976 &                    --- &                    --- &                    --- &  2.52e-16$\pm$5.75e-18 &  3.74e-16$\pm$7.79e-18 \\
BSS19 &  12.585603 &  58.116081 &  424323291174262400 &                    --- &                    --- &  1.13e-15$\pm$6.10e-17 &                    --- &                    --- \\
BSS20 &  12.703872 &  58.117638 &  424416268630796672 &                    --- &                    --- &  6.26e-17$\pm$2.11e-17 &                    --- &                    --- \\
BSS21 &  12.916705 &  58.156035 &  424414928600979712 &                    --- &                    --- &                    --- &                    --- &                    --- \\
BSS22 &  12.792388 &  58.127478 &  424416436124233344 &                    --- &                    --- &  4.66e-17$\pm$2.28e-17 &                    --- &                    --- \\
BSS23 &  12.755911 &  58.107444 &  424416096832121088 &                    --- &                    --- &                    --- &                    --- &                    --- \\
BSS24 &  12.600595 &  58.148060 &  424417265062966656 &                    --- &                    --- &                    --- &                    --- &                    --- \\
BSS25 &  12.792363 &  58.189219 &  424418188471143040 &                    --- &                    --- &                    --- &  2.71e-16$\pm$3.56e-18 &  3.99e-16$\pm$3.69e-18 \\
BSS26 &  12.744375 &  58.209324 &  424418394634254720 &                    --- &                    --- &  3.23e-16$\pm$5.68e-17 &  5.17e-16$\pm$8.32e-18 &  7.68e-16$\pm$1.01e-17 \\
BSS27 &  12.669486 &  58.259244 &  424419395366848640 &                    --- &                    --- &                    --- &                    --- &                    --- \\
BSS28 &  12.751046 &  58.197227 &  424418360269944320 &                    --- &                    --- &  1.73e-16$\pm$4.38e-17 &    4.46e-16$\pm$0.e+00 &    6.75e-16$\pm$0.e+00 \\
BSS29 &  12.651172 &  58.264862 &  424419601519284096 &                    --- &                    --- &  3.58e-16$\pm$4.71e-17 &                    --- &                    --- \\
BSS30 &  12.944214 &  58.230622 &  424421456951191040 &                    --- &                    --- &                    --- &                    --- &                    --- \\
BSS31 &  12.741104 &  58.182103 &  424416887106023168 &                    --- &                    --- &                    --- &  3.32e-16$\pm$5.35e-18 &  4.67e-16$\pm$6.12e-18 \\
BSS32 &  12.736065 &  58.182505 &  424416887106023040 &                    --- &                    --- &                    --- &  2.64e-16$\pm$6.02e-18 &  4.07e-16$\pm$8.46e-18 \\
BSS33 &  12.759065 &  58.182083 &  424416887106024704 &                    --- &                    --- &  7.26e-17$\pm$3.01e-17 &  2.12e-16$\pm$3.40e-18 &  3.14e-16$\pm$4.12e-18 \\
BSS34 &  12.516954 &  58.129105 &  425073944081350784 &                    --- &                    --- &                    --- &                    --- &                    --- \\
BSS35 &  12.507227 &  58.180810 &  425074188899601024 &                    --- &                    --- &                    --- &                    --- &                    --- \\
BSS36 &  12.551578 &  58.191653 &  425168094064999808 &                    --- &                    --- &  8.47e-17$\pm$2.75e-17 &                    --- &                    --- \\
BSS37 &  12.533187 &  58.207971 &  425168162784475008 &                    --- &                    --- &                    --- &                    --- &                    --- \\
BSS38 &  12.530339 &  58.248740 &  425168609461065344 &                    --- &                    --- &                    --- &                    --- &                    --- \\
BSS39 &  12.532937 &  58.265730 &  425168918698706688 &                    --- &                    --- &                    --- &                    --- &                    --- \\
\bottomrule
\end{tabular}
\end{table}
\end{landscape}

\newpage
\renewcommand{\thetable}{\arabic{table} (Continued...)}
\addtocounter{table}{-1}

\begin{landscape}
\begin{table}
\caption{}
\footnotesize
\begin{tabular}{lllllllll}
\toprule
Name   &         CAHA.V$\pm$err &         CAHA.R$\pm$err &      GAIA3.Gbp$\pm$err &        GAIA3.G$\pm$err &      GAIA3.Grp$\pm$err &          PS1.r$\pm$err &          PS1.g$\pm$err &          PS1.i$\pm$err \\
\midrule
BSS1  &   1.04e-15$\pm$9.6e-18 &  7.48e-16$\pm$1.56e-17 &  1.05e-15$\pm$3.77e-18 &  9.17e-16$\pm$2.42e-18 &  6.33e-16$\pm$2.98e-18 &  8.98e-16$\pm$2.18e-18 &  1.17e-15$\pm$6.58e-18 &  6.57e-16$\pm$3.91e-18 \\
BSS2  &                    --- &                    --- &  3.31e-15$\pm$9.33e-18 &    2.73e-15$\pm$7.e-18 &   1.76e-15$\pm$6.5e-18 &  2.63e-15$\pm$4.42e-18 &  3.73e-15$\pm$3.04e-18 &  1.81e-15$\pm$5.46e-18 \\
BSS3  &  1.66e-15$\pm$1.53e-17 &  1.17e-15$\pm$1.54e-17 &  1.74e-15$\pm$1.18e-17 &  1.44e-15$\pm$3.74e-18 &  1.14e-15$\pm$8.38e-18 &  1.47e-15$\pm$1.16e-17 &  1.82e-15$\pm$6.67e-18 &  1.10e-15$\pm$1.27e-17 \\
BSS4  &    9.37e-16$\pm$0.e+00 &  6.64e-16$\pm$6.14e-18 &  9.13e-16$\pm$4.15e-18 &  7.77e-16$\pm$2.13e-18 &  5.48e-16$\pm$3.35e-18 &   7.69e-16$\pm$4.1e-18 &  9.92e-16$\pm$5.18e-18 &  5.56e-16$\pm$3.92e-18 \\
BSS5  &                    --- &                    --- &  3.95e-16$\pm$8.03e-18 &  3.82e-16$\pm$2.38e-18 &  2.96e-16$\pm$5.61e-18 &  4.26e-16$\pm$1.86e-18 &  4.86e-16$\pm$1.02e-17 &  3.60e-16$\pm$9.59e-19 \\
BSS6  &                    --- &                    --- &  4.35e-16$\pm$2.74e-18 &  4.31e-16$\pm$1.12e-18 &  3.47e-16$\pm$1.73e-18 &  4.34e-16$\pm$1.30e-18 &  4.47e-16$\pm$6.11e-19 &  3.57e-16$\pm$1.66e-18 \\
BSS7  &                    --- &                    --- &  2.22e-15$\pm$6.52e-18 &  1.81e-15$\pm$4.63e-18 &  1.15e-15$\pm$4.43e-18 &  1.75e-15$\pm$2.87e-18 &  2.52e-15$\pm$4.96e-18 &   1.19e-15$\pm$1.7e-18 \\
BSS8  &                    --- &                    --- &  2.17e-15$\pm$9.38e-18 &  1.78e-15$\pm$5.63e-18 &  1.44e-15$\pm$6.71e-18 &  1.94e-15$\pm$4.52e-18 &  2.38e-15$\pm$2.27e-17 &  1.51e-15$\pm$6.08e-18 \\
BSS9  &                    --- &                    --- &  1.24e-15$\pm$3.85e-18 &  1.10e-15$\pm$2.83e-18 &  7.85e-16$\pm$2.92e-18 &  1.09e-15$\pm$2.63e-18 &  1.36e-15$\pm$5.10e-18 &  8.17e-16$\pm$1.91e-18 \\
BSS10 &                    --- &                    --- &  5.81e-16$\pm$2.82e-18 &  5.97e-16$\pm$1.54e-18 &  4.86e-16$\pm$2.03e-18 &  5.96e-16$\pm$1.29e-18 &  5.98e-16$\pm$1.66e-18 &  5.06e-16$\pm$1.76e-18 \\
BSS11 &                    --- &                    --- &  2.71e-15$\pm$8.32e-18 &  2.54e-15$\pm$6.92e-18 &  2.06e-15$\pm$7.59e-18 &  2.62e-15$\pm$3.41e-18 &  2.79e-15$\pm$1.59e-17 &  2.10e-15$\pm$4.01e-18 \\
BSS12 &  4.57e-16$\pm$8.49e-18 &    3.3e-16$\pm$8.7e-18 &  5.41e-16$\pm$5.55e-18 &  4.12e-16$\pm$1.11e-18 &   3.8e-16$\pm$5.22e-18 &  4.71e-16$\pm$2.73e-18 &   5.4e-16$\pm$1.11e-17 &  3.12e-16$\pm$3.80e-19 \\
BSS13 &                    --- &                    --- &  5.22e-16$\pm$2.54e-18 &   5.1e-16$\pm$1.32e-18 &  3.91e-16$\pm$1.81e-18 &  5.17e-16$\pm$1.98e-18 &  5.58e-16$\pm$1.13e-18 &   4.1e-16$\pm$2.38e-18 \\
BSS14 &                    --- &                    --- &  5.58e-16$\pm$2.88e-18 &   5.40e-16$\pm$1.4e-18 &  4.18e-16$\pm$1.93e-18 &  5.41e-16$\pm$1.41e-18 &  5.94e-16$\pm$3.12e-18 &  4.32e-16$\pm$1.15e-18 \\
BSS15 &                    --- &                    --- &  5.08e-16$\pm$2.38e-18 &  4.93e-16$\pm$1.28e-18 &  3.91e-16$\pm$1.87e-18 &  4.93e-16$\pm$4.97e-18 &  5.24e-16$\pm$2.41e-18 &  3.99e-16$\pm$9.95e-19 \\
BSS16 &                    --- &                    --- &   3.89e-16$\pm$2.6e-18 &  3.80e-16$\pm$1.00e-18 &  2.93e-16$\pm$1.67e-18 &  3.82e-16$\pm$1.94e-18 &  4.24e-16$\pm$1.89e-18 &  3.04e-16$\pm$1.66e-18 \\
BSS17 &                    --- &                    --- &  6.09e-16$\pm$2.68e-18 &  5.79e-16$\pm$1.49e-18 &    4.37e-16$\pm$2.e-18 &   5.8e-16$\pm$1.76e-18 &  6.61e-16$\pm$2.81e-18 &  4.55e-16$\pm$2.08e-18 \\
BSS18 &    4.32e-16$\pm$4.e-18 &  3.48e-16$\pm$7.25e-18 &  4.26e-16$\pm$2.36e-18 &  4.14e-16$\pm$1.11e-18 &  3.21e-16$\pm$1.65e-18 &  4.14e-16$\pm$1.49e-18 &  4.59e-16$\pm$1.32e-18 &  3.29e-16$\pm$2.25e-18 \\
BSS19 &                    --- &                    --- &  1.11e-15$\pm$1.67e-17 &  9.60e-16$\pm$2.47e-18 &  7.52e-16$\pm$5.66e-18 &  1.11e-15$\pm$6.75e-17 &  1.09e-15$\pm$4.76e-18 &  7.48e-16$\pm$3.34e-18 \\
BSS20 &                    --- &                    --- &  3.91e-16$\pm$2.34e-18 &  3.95e-16$\pm$1.03e-18 &  3.22e-16$\pm$1.65e-18 &  3.92e-16$\pm$1.55e-18 &  4.03e-16$\pm$1.12e-18 &  3.34e-16$\pm$1.13e-18 \\
BSS21 &                    --- &                    --- &   7.4e-16$\pm$3.48e-18 &  7.54e-16$\pm$1.94e-18 &  6.12e-16$\pm$2.69e-18 &  7.54e-16$\pm$2.82e-18 &  7.69e-16$\pm$1.61e-18 &  6.36e-16$\pm$1.77e-18 \\
BSS22 &                    --- &                    --- &  7.48e-16$\pm$3.48e-18 &  6.97e-16$\pm$1.83e-18 &  6.42e-16$\pm$2.76e-18 &  7.62e-16$\pm$2.52e-18 &  7.44e-16$\pm$1.84e-18 &  6.41e-16$\pm$2.33e-18 \\
BSS23 &                    --- &                    --- &  6.58e-16$\pm$3.64e-18 &  6.49e-16$\pm$1.68e-18 &  5.25e-16$\pm$2.48e-18 &  6.58e-16$\pm$3.49e-18 &  6.84e-16$\pm$1.84e-18 &  5.47e-16$\pm$1.15e-18 \\
BSS24 &                    --- &                    --- &  6.26e-16$\pm$2.97e-18 &  5.66e-16$\pm$1.46e-18 &  4.01e-16$\pm$1.65e-18 &  5.70e-16$\pm$1.94e-18 &    6.87e-16$\pm$3.e-18 &  4.21e-16$\pm$1.53e-18 \\
BSS25 &    5.25e-16$\pm$0.e+00 &  4.42e-16$\pm$4.09e-18 &  4.83e-16$\pm$2.45e-18 &  4.99e-16$\pm$1.29e-18 &  4.11e-16$\pm$1.73e-18 &  4.99e-16$\pm$1.66e-18 &  4.92e-16$\pm$2.24e-18 &  4.20e-16$\pm$1.15e-18 \\
BSS26 &  7.72e-16$\pm$7.15e-18 &  5.47e-16$\pm$1.14e-17 &  7.67e-16$\pm$3.92e-18 &   6.56e-16$\pm$1.7e-18 &  4.77e-16$\pm$2.45e-18 &  6.63e-16$\pm$1.22e-18 &  8.35e-16$\pm$3.78e-18 &  4.83e-16$\pm$2.19e-18 \\
BSS27 &                    --- &                    --- &  3.13e-16$\pm$2.65e-18 &  3.22e-16$\pm$8.48e-19 &  2.68e-16$\pm$1.53e-18 &  3.21e-16$\pm$1.05e-18 &  3.14e-16$\pm$1.08e-18 &  2.76e-16$\pm$1.51e-18 \\
BSS28 &    7.04e-16$\pm$0.e+00 &    5.13e-16$\pm$0.e+00 &  7.09e-16$\pm$3.49e-18 &  6.11e-16$\pm$1.58e-18 &  4.58e-16$\pm$3.21e-18 &  6.28e-16$\pm$3.32e-18 &  7.66e-16$\pm$6.66e-18 &  4.64e-16$\pm$4.53e-18 \\
BSS29 &                    --- &                    --- &  4.93e-16$\pm$1.33e-17 &  3.24e-16$\pm$1.01e-18 &  3.96e-16$\pm$5.41e-18 &                    --- &                    --- &  3.53e-16$\pm$1.56e-18 \\
BSS30 &                    --- &                    --- &  3.99e-16$\pm$2.37e-18 &  4.11e-16$\pm$1.07e-18 &  3.42e-16$\pm$1.91e-18 &  4.11e-16$\pm$1.73e-18 &  4.04e-16$\pm$1.69e-18 &  3.53e-16$\pm$2.26e-18 \\
BSS31 &  4.40e-16$\pm$4.07e-18 &  3.03e-16$\pm$6.31e-18 &  4.42e-16$\pm$1.99e-18 &  3.77e-16$\pm$9.87e-19 &  2.48e-16$\pm$1.33e-18 &  3.69e-16$\pm$1.07e-18 &    5.e-16$\pm$2.37e-18 &   2.6e-16$\pm$7.87e-19 \\
BSS32 &  5.01e-16$\pm$4.64e-18 &  4.19e-16$\pm$8.71e-18 &  4.91e-16$\pm$6.23e-18 &  4.95e-16$\pm$2.30e-18 &  3.89e-16$\pm$4.94e-18 &  5.35e-16$\pm$3.38e-18 &  4.50e-16$\pm$3.41e-18 &  3.73e-16$\pm$1.09e-17 \\
BSS33 &   3.7e-16$\pm$3.42e-18 &  2.92e-16$\pm$6.08e-18 &  3.58e-16$\pm$1.94e-18 &  3.44e-16$\pm$9.05e-19 &   2.7e-16$\pm$1.42e-18 &  3.48e-16$\pm$1.96e-18 &  3.74e-16$\pm$1.76e-18 &  2.77e-16$\pm$1.17e-18 \\
BSS34 &                    --- &                    --- &  1.14e-15$\pm$4.32e-18 &  1.16e-15$\pm$2.97e-18 &  9.36e-16$\pm$3.78e-18 &  1.16e-15$\pm$2.23e-18 &   1.2e-15$\pm$4.82e-18 &  9.64e-16$\pm$3.07e-18 \\
BSS35 &                    --- &                    --- &   4.02e-16$\pm$2.9e-18 &  4.09e-16$\pm$1.07e-18 &  3.32e-16$\pm$1.81e-18 &  4.04e-16$\pm$2.08e-18 &  4.21e-16$\pm$2.32e-18 &  3.41e-16$\pm$1.22e-18 \\
BSS36 &                    --- &                    --- &  4.58e-16$\pm$2.33e-18 &  4.61e-16$\pm$1.19e-18 &   3.65e-16$\pm$1.9e-18 &  4.63e-16$\pm$1.08e-18 &  4.76e-16$\pm$1.54e-18 &  3.74e-16$\pm$1.76e-18 \\
BSS37 &                    --- &                    --- &  3.98e-16$\pm$2.75e-18 &  4.02e-16$\pm$1.04e-18 &  3.25e-16$\pm$1.65e-18 &   4.04e-16$\pm$2.1e-18 &  4.04e-16$\pm$2.03e-18 &  3.36e-16$\pm$3.62e-18 \\
BSS38 &                    --- &                    --- &  5.62e-16$\pm$2.29e-18 &  5.72e-16$\pm$1.48e-18 &  4.63e-16$\pm$2.08e-18 &  5.77e-16$\pm$2.91e-18 &  5.80e-16$\pm$2.53e-18 &  4.79e-16$\pm$2.68e-18 \\
BSS39 &                    --- &                    --- &  8.86e-16$\pm$3.28e-18 &  8.86e-16$\pm$2.27e-18 &  7.11e-16$\pm$2.90e-18 &  9.05e-16$\pm$5.27e-18 &  9.20e-16$\pm$2.39e-18 &  7.36e-16$\pm$1.62e-18 \\
\bottomrule
\end{tabular}

\end{table}
\end{landscape}

\newpage
\renewcommand{\thetable}{\arabic{table} (Continued...)}
\addtocounter{table}{-1}

\begin{landscape}
\begin{table}
\caption{}
\footnotesize
\begin{tabular}{llllll}
\toprule
Name   &          PS1.z$\pm$err &          PS1.y$\pm$err &        2MASS.J$\pm$err &        2MASS.H$\pm$err &       2MASS.Ks$\pm$err \\
\midrule
BSS1  &  5.25e-16$\pm$1.08e-18 &  4.74e-16$\pm$2.72e-18 &  2.62e-16$\pm$2.54e-17 &  1.13e-16$\pm$1.34e-17 &  5.08e-17$\pm$4.92e-18 \\
BSS2  &  1.39e-15$\pm$4.31e-18 &  1.22e-15$\pm$3.58e-18 &  6.45e-16$\pm$1.90e-17 &  2.62e-16$\pm$9.67e-18 &  1.11e-16$\pm$5.71e-18 \\
BSS3  &  9.14e-16$\pm$1.53e-17 &  8.26e-16$\pm$5.47e-18 &  5.06e-16$\pm$2.24e-17 &  2.38e-16$\pm$1.18e-17 &  9.34e-17$\pm$5.51e-18 \\
BSS4  &   4.4e-16$\pm$3.34e-18 &  3.93e-16$\pm$3.24e-18 &  2.49e-16$\pm$2.09e-17 &  1.23e-16$\pm$1.08e-17 &  5.51e-17$\pm$4.98e-18 \\
BSS5  &  2.43e-16$\pm$5.71e-18 &  2.58e-16$\pm$1.69e-18 &  1.72e-16$\pm$8.39e-18 &  8.52e-17$\pm$8.16e-18 &   3.49e-17$\pm$4.5e-18 \\
BSS6  &  2.99e-16$\pm$9.55e-19 &  2.69e-16$\pm$2.31e-18 &  1.68e-16$\pm$8.98e-18 &  8.63e-17$\pm$7.15e-18 &  2.51e-17$\pm$4.62e-18 \\
BSS7  &  9.12e-16$\pm$3.76e-18 &  8.00e-16$\pm$3.15e-18 &  4.13e-16$\pm$1.10e-17 &  1.62e-16$\pm$8.08e-18 &  7.57e-17$\pm$4.60e-18 \\
BSS8  &  1.18e-15$\pm$9.73e-18 &  1.04e-15$\pm$7.33e-18 &  6.38e-16$\pm$1.82e-17 &  2.81e-16$\pm$1.14e-17 &  1.35e-16$\pm$6.98e-18 \\
BSS9  &  6.41e-16$\pm$1.70e-18 &  5.95e-16$\pm$4.39e-18 &  3.28e-16$\pm$1.18e-17 &  1.22e-16$\pm$7.77e-18 &  6.14e-17$\pm$4.64e-18 \\
BSS10 &  4.28e-16$\pm$1.07e-18 &  3.83e-16$\pm$2.33e-18 &  2.35e-16$\pm$9.74e-18 &   1.1e-16$\pm$7.49e-18 &  4.01e-17$\pm$4.43e-18 \\
BSS11 &  1.71e-15$\pm$9.50e-19 &  1.49e-15$\pm$4.28e-18 &  8.75e-16$\pm$2.74e-17 &  4.44e-16$\pm$1.55e-17 &  1.79e-16$\pm$6.59e-18 \\
BSS12 &  2.62e-16$\pm$7.86e-18 &  2.29e-16$\pm$3.82e-18 &                    --- &                    --- &                    --- \\
BSS13 &  3.31e-16$\pm$1.17e-18 &  2.94e-16$\pm$1.54e-18 &  1.90e-16$\pm$8.77e-18 &  7.19e-17$\pm$8.28e-18 &  3.57e-17$\pm$4.53e-18 \\
BSS14 &  3.52e-16$\pm$1.72e-18 &  3.15e-16$\pm$9.87e-19 &  1.81e-16$\pm$9.68e-18 &   9.1e-17$\pm$8.05e-18 &  3.39e-17$\pm$4.47e-18 \\
BSS15 &  3.27e-16$\pm$8.77e-19 &    3.e-16$\pm$2.43e-18 &  1.76e-16$\pm$1.34e-17 &  8.93e-17$\pm$1.00e-17 &  3.82e-17$\pm$4.71e-18 \\
BSS16 &  2.48e-16$\pm$1.38e-18 &  2.17e-16$\pm$1.34e-18 &  1.22e-16$\pm$8.34e-18 &  4.58e-17$\pm$7.26e-18 &  3.06e-17$\pm$4.23e-18 \\
BSS17 &  3.69e-16$\pm$9.43e-19 &  3.35e-16$\pm$1.67e-18 &  1.95e-16$\pm$9.35e-18 &  7.14e-17$\pm$8.22e-18 &   2.53e-17$\pm$4.2e-18 \\
BSS18 &  2.72e-16$\pm$1.06e-18 &  2.48e-16$\pm$4.72e-18 &  1.40e-16$\pm$1.03e-17 &  7.21e-17$\pm$7.51e-18 &  2.63e-17$\pm$4.23e-18 \\
BSS19 &  6.05e-16$\pm$2.66e-18 &  5.45e-16$\pm$5.40e-18 &                    --- &                    --- &                    --- \\
BSS20 &  2.76e-16$\pm$1.32e-18 &  2.49e-16$\pm$1.19e-18 &  1.40e-16$\pm$8.54e-18 &  6.93e-17$\pm$7.35e-18 &  3.35e-17$\pm$4.38e-18 \\
BSS21 &  5.32e-16$\pm$2.70e-18 &   4.8e-16$\pm$2.17e-18 &  2.52e-16$\pm$1.02e-17 &  1.42e-16$\pm$7.47e-18 &  5.78e-17$\pm$4.63e-18 \\
BSS22 &  5.30e-16$\pm$3.77e-18 &  4.75e-16$\pm$3.28e-18 &  2.99e-16$\pm$1.26e-17 &  1.65e-16$\pm$8.21e-18 &  6.69e-17$\pm$4.81e-18 \\
BSS23 &  4.51e-16$\pm$1.68e-18 &  4.12e-16$\pm$1.62e-18 &  2.41e-16$\pm$1.84e-17 &  1.08e-16$\pm$1.24e-17 &  3.85e-17$\pm$4.39e-18 \\
BSS24 &  3.34e-16$\pm$1.49e-18 &  2.97e-16$\pm$1.90e-18 &  1.63e-16$\pm$1.14e-17 &  8.01e-17$\pm$8.78e-18 &  3.33e-17$\pm$4.36e-18 \\
BSS25 &  3.57e-16$\pm$1.07e-18 &  3.24e-16$\pm$3.41e-18 &  1.80e-16$\pm$9.14e-18 &  9.28e-17$\pm$7.43e-18 &  3.86e-17$\pm$4.62e-18 \\
BSS26 &  3.92e-16$\pm$2.16e-18 &  3.78e-16$\pm$2.60e-18 &  2.06e-16$\pm$2.72e-17 &  1.02e-16$\pm$1.42e-17 &    2.97e-16$\pm$0.e+00 \\
BSS27 &  2.34e-16$\pm$8.07e-19 &  2.15e-16$\pm$2.34e-18 &  1.29e-16$\pm$9.73e-18 &   5.7e-17$\pm$8.24e-18 &  2.50e-17$\pm$4.08e-18 \\
BSS28 &  3.72e-16$\pm$2.29e-18 &  3.35e-16$\pm$2.27e-18 &  1.84e-16$\pm$1.77e-17 &    8.e-17$\pm$1.08e-17 &  3.44e-17$\pm$5.70e-18 \\
BSS29 &  2.83e-16$\pm$3.36e-19 &                    --- &                    --- &                    --- &                    --- \\
BSS30 &  2.97e-16$\pm$2.38e-18 &   2.73e-16$\pm$2.7e-18 &  1.75e-16$\pm$1.01e-17 &  7.47e-17$\pm$7.36e-18 &  4.95e-17$\pm$4.47e-18 \\
BSS31 &  1.97e-16$\pm$8.58e-19 &  1.76e-16$\pm$1.28e-18 &  8.13e-17$\pm$7.64e-18 &    5.44e-17$\pm$0.e+00 &    1.57e-17$\pm$0.e+00 \\
BSS32 &  3.29e-16$\pm$7.91e-18 &  2.68e-16$\pm$6.06e-18 &   1.60e-16$\pm$1.2e-17 &  8.60e-17$\pm$1.11e-17 &  3.44e-17$\pm$4.47e-18 \\
BSS33 &   2.3e-16$\pm$9.57e-19 &   2.1e-16$\pm$2.92e-18 &  1.16e-16$\pm$1.36e-17 &  5.84e-17$\pm$9.30e-18 &  3.00e-17$\pm$4.79e-18 \\
BSS34 &  8.05e-16$\pm$1.36e-18 &  7.30e-16$\pm$2.53e-18 &   4.40e-16$\pm$1.3e-17 &  2.05e-16$\pm$8.11e-18 &  9.78e-17$\pm$4.96e-18 \\
BSS35 &  2.88e-16$\pm$9.61e-19 &   2.6e-16$\pm$3.76e-18 &  1.80e-16$\pm$9.80e-18 &  7.33e-17$\pm$6.75e-18 &   3.7e-17$\pm$4.26e-18 \\
BSS36 &  3.17e-16$\pm$4.78e-19 &  2.78e-16$\pm$3.62e-18 &  1.68e-16$\pm$9.89e-18 &  8.51e-17$\pm$6.97e-18 &  3.27e-17$\pm$4.31e-18 \\
BSS37 &   2.8e-16$\pm$2.77e-18 &  2.51e-16$\pm$2.45e-18 &  1.60e-16$\pm$9.29e-18 &  7.96e-17$\pm$6.89e-18 &  3.00e-17$\pm$4.59e-18 \\
BSS38 &  3.99e-16$\pm$2.09e-18 &  3.57e-16$\pm$1.09e-18 &  2.24e-16$\pm$1.07e-17 &  1.07e-16$\pm$7.31e-18 &  4.86e-17$\pm$4.70e-18 \\
BSS39 &  6.08e-16$\pm$1.66e-18 &  5.54e-16$\pm$2.85e-18 &  3.34e-16$\pm$1.23e-17 &  1.46e-16$\pm$8.77e-18 &  6.36e-17$\pm$4.69e-18 \\
\bottomrule
\end{tabular}

\end{table}
\end{landscape}

\renewcommand{\thetable}{\arabic{table}}
\end{document}